\documentclass[10pt,journal]{IEEEtranTCOM}

\usepackage[english]{babel}
\usepackage[usenames]{color}
\usepackage[cp1250]{inputenc}
\usepackage{amsfonts}
\usepackage{amsthm}
\usepackage{graphicx}
\usepackage{epsfig}
\usepackage{mathrsfs}
\usepackage{amsmath}
\usepackage{algorithm}
\usepackage{algorithmic}

\theoremstyle{plain}

\begin{document}
\newcommand{\bea}{\begin{eqnarray}}
\newcommand{\eea}{\end{eqnarray}}
\newcommand{\be}{\begin{equation}}
\newcommand{\ee}{\end{equation}}
\newcommand{\beas}{\begin{eqnarray*}}
\newcommand{\eeas}{\end{eqnarray*}}
\newcommand{\bs}{\backslash}
\newcommand{\bc}{\begin{center}}
\newcommand{\ec}{\end{center}}
\def\SC {\mathscr{C}}

\title{Joint error correction enhancement \\of the fountain codes concept}
\author{\IEEEauthorblockN{Jarek Duda}\\
\IEEEauthorblockA{Jagiellonian University,
Golebia 24, 31-007 Krakow, Poland,
Email: \emph{dudajar@gmail.com}}}
\maketitle

\begin{abstract}
Fountain codes like LT or Raptor codes, also known as rateless erasure codes, allow to encode a message as some number of packets, such that any large enough subset of these packets is sufficient to fully reconstruct the message. It requires undamaged packets, while the packets which were not lost are usually damaged in real scenarios. Hence, an additional error correction layer is often required: adding some level of redundancy to each packet to be able to repair eventual damages. This approach requires a priori knowledge of the final damage level of every packet - insufficient redundancy leads to packet loss, overprotection means suboptimal channel rate. However, the sender may have inaccurate or even no a priori information about the final damage levels, for example in applications like broadcasting, degradation of a storage medium or damage of picture watermarking.

Joint Reconstruction Codes (JRC) setting is introduced and discussed in this paper for the purpose of removing the need of a priori knowledge of damage level and sub-optimality caused by overprotection and discarding underprotected packets. It is obtained by combining both processes: reconstruction from multiple packets and forward error correction. Intuitively, instead of adding artificial redundancy to each packet, the packets are prepared to be simultaneously payload and redundancy. The decoder combines the resultant informational content of all received packets accordingly to their actual noise level, which can be estimated a posteriori individually for each packet. Assuming binary symmetric channel (BSC) of $\epsilon$ bit-flip probability, every potentially damaged bit carries $R_0(\epsilon)=1-h_1(\epsilon)$ bits of information, where $h_1$ is the Shannon entropy. The minimal requirement to fully reconstruct the message is that the sum of rate $R_0(\epsilon)$ over all bits is at least the size of the message. We will discuss sequential decoding for the reconstruction purpose, which allows to work close to the theoretical limit. Analysis and tests of the accompanied implementation show that the statistical behavior of constructed tree can be approximated by Pareto distribution with coefficient $c$ such that the sum of $R_c(\epsilon)=1-h_{1/(1+c)}(\epsilon)$ is the size of the message, where $h_u(\epsilon)=(\epsilon^u+(1-\epsilon)^u)/(1-u)$ is $u$-th Renyi entropy.
\end{abstract}
\textbf{Keywords:} fountain codes, error correction, sequential decoding, Renyi entropy, broadcasting
\section{Introduction}
\begin{figure}[t!]
    \centering
        \includegraphics[width=8cm]{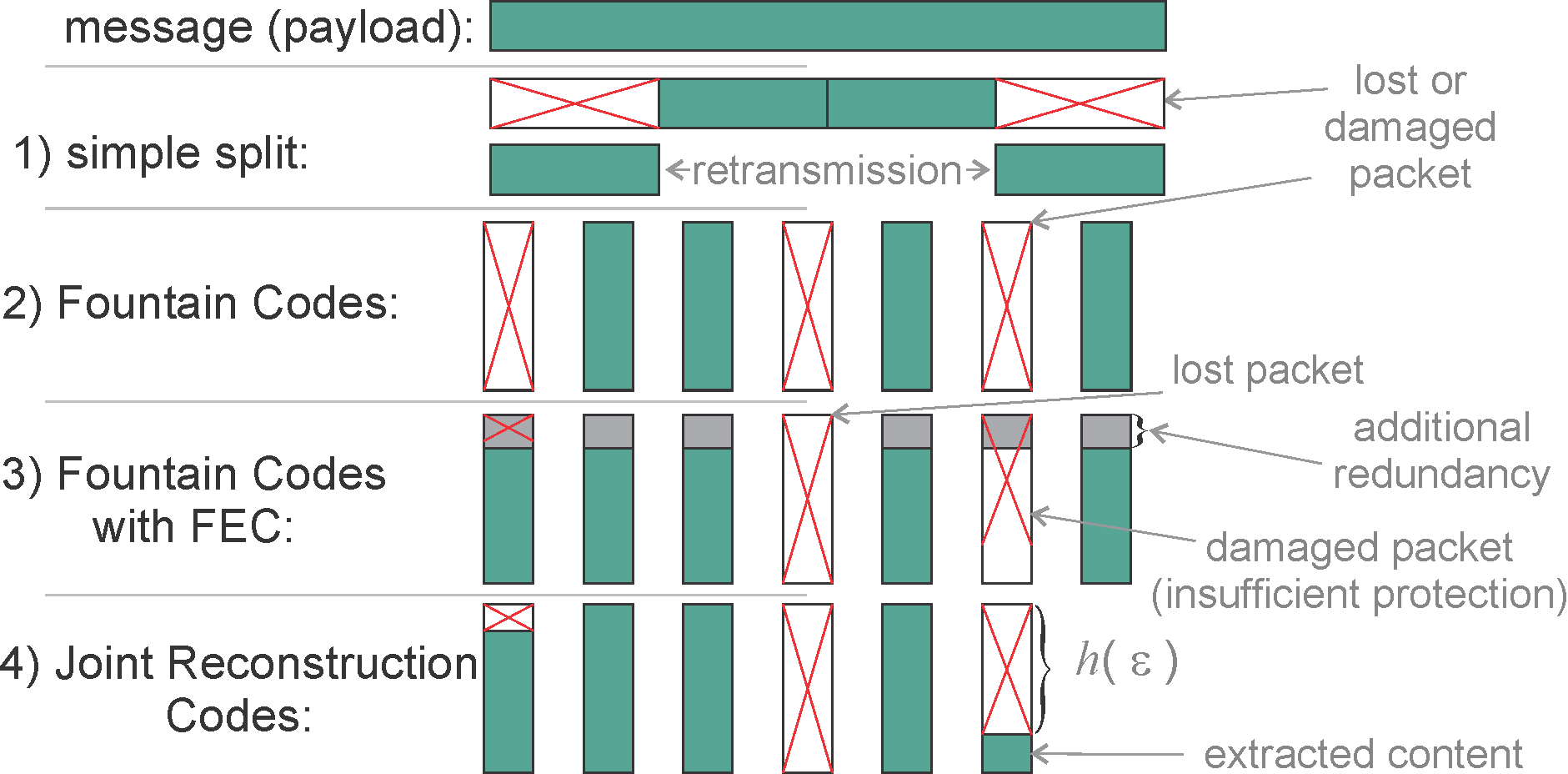}
        \caption{Schematic diagram of different possibilities of splitting the informational content of the message. 1) Simple split requires all undamaged packets. 2) For fountain codes, any large enough subset of undamaged packets is sufficient. 3) While the set of packets is already redundant, we can add additional redundancy to separately protect every packet - some turn out overprotected, some insufficiently. 4) JRC allows to extract the actual informational content (green area) from all obtained packets. Reconstruction requires to gather a sufficient amount of informational content: such that the sum of $1-h(\epsilon)$ over all bits exceeds the size of the message, where $\epsilon$ is the bit-flip probability of assumed binary symmetric channel (BSC).}
        \label{joint}
\end{figure}

Many communication settings require dividing a potentially large message (payload) into relatively small blocks of data, which will be referred as packets. These packets are often damaged or even lost on the way. Both issues could be handled by allowing for retransmission of uncertain or missing parts of data. However, it would require costly bidirectional communication, waiting for packets before classifying them as lost (automatic repeat request after a timeout). Hence, there is a strong need for methods allowing to reduce or completely eliminate the feedback, schematically presented in Fig. \ref{joint}.

The issue of missing packets can be handled by various erasure channel methods, for example by the fact that coefficients of degree $d$ polynomial can be obtained from values in any $d+1$ points in classic Reed-Solomon codes~\cite{RS}. A faster way, pioneered in LT codes by  Michael Luby~\cite{Luby} in 2002, is constructing the packets as XOR of some subsets of data blocks, usually chosen in a random way. This approach is currently referred as fountain codes(FC) or rateless erasure codes and was improved to linear encoding and decoding time in Raptor Codes~\cite{raptor}. There are considered various applications of FC like relay networks~\cite{relay}, sensor networks~\cite{sensor}, radio networks~\cite{radio}, networked storage~\cite{storage} or watermaring~\cite{water}.

If a packet was not lost, it is often received damaged. Forward error correction (FEC) layer is used to handle this issue without retransmission: the sender attaches some redundancy, which will be used by decoder to repair eventual damages. The applied redundancy level determines a statistical threshold for damage, deciding if a given packet can be repaired. Fountain codes do not tolerate damaged packets, so those exceeding the damage threshold need to be discarded - their entire informational content is being lost. As this threshold has a statistical nature, this all-or-nothing means inefficient use of the channel: some packets turn out to be overprotected - have used an unnecessarily high amount of redundancy, some packets insufficiently - loosing all the content.

Additionally, in many scenarios the sender has inaccurate or no a priori knowledge of the final damage level, hence some overprotection needs to be used, and some of packets are still being lost. For example
\begin{itemize}
  \item{broadcasting: while every packet (receiver) may have a different individual damage level, the sender needs to prepare universal packets,}
  \item{protecting storage media: the final damage level depends on age, conditions and random accidents,}
  \item{sending a packet through a network: there is often no knowledge of the route it will take,}
  \item{embedding a (robust) watermaring: there is no knowledge of the final capturing conditions,}
  \item{in some scenarios the noise level can vary too quickly for adaptive change of redundancy level, for example in acoustic or radio communication (e.g. underwater).}
\end{itemize}
This damage level can be usually estimated a posteriori - correspondingly: basing on the conditions for individual packets, history of given storage medium, on the actual route of a given packet, on the parameters of captured image, on more recent evaluation of environmental parameters. Moreover, the applied redundancy could allow to estimate the actual damage level from the received packets. Therefore, it would be beneficial to allow to shift the need of knowledge of damage level from a priori to a posteriori. \\

Joint Reconstruction Codes (JRC) setting is introduced, analyzed and discussed for this purpose. While offering the same rates, they require only a posteriori knowledge of the damage level thanks to combining both processes: of error correction and of reconstruction from multiple packets. The sender prepares packets to be treated simultaneously as the message (payload) and FEC redundancy. The decoder searches the space of promising candidates for the message sequence, to get the best agreement with all the received packets accordingly to their a posteriori damage level - individual level of trust for each packet.
\begin{figure}[t!]
    \centering
        \includegraphics[width=8cm]{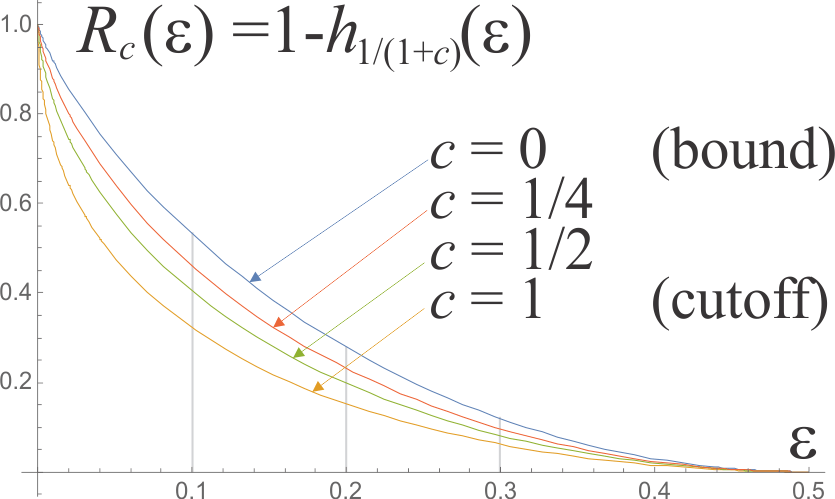}
        \caption{Rate: informational content per bit assuming binary symmetric channel (BSC). Renyi entropy $h_u(\epsilon)=\lg(\epsilon^u+(1-\epsilon)^u)/(1-u)$ becomes Shannon entropy for $u\to 1$ ($c\to 0$). The $c=0$ rate is the theoretical bound. The remaining cases allow to find Pareto coefficient $c$ describing statistical behavior of the correction process: such that sum of $R_{c}(\epsilon)$ over all bits is the size of the message.}
        \label{entr}
\end{figure}

For simplicity we will only consider the binary symmetric channel (BSC) here: that every bit has independent $\epsilon$ probability of being flipped. Theoretical rate limit in this case is $R_0(\epsilon):=1-h(\epsilon)$, where $h(p)=-p\lg(p)-(1-p)\lg(1-p)$ is the Shannon entropy. Therefore, the minimal requirement for reconstructing a message is that the sum of $1-h(\epsilon)$ over all received bits is at least the size of the message. It is schematically presented in Fig. \ref{joint} using green bars containing $1-h(\epsilon)$ of size of the packet - the minimal requirement for reconstruction is that the green bars sum to the length of the message.

Practical error correction methods usually work below this theoretical rate. We will discuss implemented enhancement of sequential decoding, for which there is known so called cutoff rate~\cite{cutoff}. This article uses a generalized family of rates described by $c\geq 0$ parameter (Pareto coefficient) as in Fig. \ref{entr}:
\be R_c(\epsilon)=1-h_{1/(1+c)} (\epsilon)\ee
where $h_u(\epsilon)=\frac{\lg(\epsilon^u+(1-\epsilon)^u)}{1-u}$ is $u$-th Renyi entropy~\cite{renyi}, $\lg\equiv \log_2$.
The $u=1$ case is naturally defined as the $u\to 1$ limit, getting the Shannon entropy:
\be h_1(\epsilon)\equiv h(\epsilon)=-\epsilon\lg(\epsilon)-(1-\epsilon)\lg(1-\epsilon).\ee
Finally, we have the two mentioned special cases:
$$ R_0(\epsilon)=1-h(\epsilon)$$ 
$$R_1(\epsilon)=2\lg\left(\sqrt\epsilon+\sqrt{1-\epsilon}\right)=\lg\left(1+2\sqrt{\epsilon(1-\epsilon)}\right)$$
$R_0$ is the theoretical rate limit for this channel. $R_1$ is the cutoff rate for sequential decoding~\cite{cutoff}: below this rate, sequential decoding of infinite message would require a finite $width$ - average number of steps per bit block. Generally, as it will be shown in Section \ref{sec4}, $c$ is the coefficient of Pareto distribution describing statistical behavior of the correction process: while increasing twice the maximal number of considered steps, probability of failure drops asymptotically $2^c$ times. Thanks of using finite data frames and some improvements to the sequential decoding: large internal state (64 bit) and bidirectional correction, much larger rates than for the $c=1$ cutoff rate can be used~\cite{cortree}, e.g. for $c=1/2$ and 1kB data frames there were still obtained complete correction in all tests for all three considered cases: 7/8, 1/2 and 3/4 rate.

Finally, the $c$ value describing correction process is such that the sum of $R_c(\epsilon)$ over all bits is the size of the message. The cost (difficulty) of correction statistically decreases with the growth of $c$. Therefore, from JRC perspective we should choose some boundary value: $c_{min}$ and wait (gather packets) until the sum of $R_{c_{min}}(\epsilon)$ over the received packets exceeds the message size. Then the decoder can try to perform the correction and eventually wait for another packet(s) to try again if it has failed to reconstruct within assumed resource limit (time and memory). For unidirectional sequential decoding (implemented) this $c_{min}$ boundary can be chosen in $\approx[0.5,1]$ range, depending on the resource limit. Bidirectional sequential decoding effectively allows to halve this value.\\

Let us briefly look at some possible applications of this general possibility of removing the need of a priori knowledge of damage level:
\begin{itemize}
  \item A basic example is broadcasting, where the sender have to use universal packets. A posteriori adaptation to the actual noise levels of individual packets (receivers) may be beneficial, and can be effectively obtained by JRC.
  \item Another basic application is protection of storage media. For example imagine a thousand of DVD copies of a movie. While time passes, all of them will degrade, finally exceeding the included protection level - making all of them useless. JRC approach would allow to still reconstruct the original content from some number of badly damaged copies, where the required number depends on damage level. As discussed in Section \ref{sec3}, a message encoded this way can be prepared for 3 levels of decoding: fast straightforward decoding from any single undamaged disc, or with error correction for a lightly damaged single disk, or with error correction for multiple heavily damaged disks.
  \item Another family of applications is improving efficiency and reducing energy consumption of various networks by replacing error correction applied by every node, with a single reconstruction+correction applied by the receiver only (of the payload - some small headers need to be frequently corrected). Decreasing energy consumption of network nodes is especially important for battery operated, e.g. sensor networks.
  \item For watermarking applications we could divide a message into small packets - the receiver should be able to reconstruct the message even if some of packets are missing (e.g. part of a picture or frames of a video) and the remaining are damaged accordingly to capturing conditions - in an unknown to the sender way.
  \item This approach can be also used when there are no missing packets (some additional optimizations can be applied in this case). For example imagine acoustic or radio communication while rapidly varying environmental conditions - the receiver can likely receive all the packets, and he usually has more information about the actual damage level than the sender, making JRC beneficial.
\end{itemize}

The situation with asymmetric sender-receiver knowledge is considered for example in the Kuznetsov-Tsybakov problem~\cite{KT}, where only the sender knows the pattern of fixed bits of the channel. In contrast, we will assume here that the receiver have more information about the noise levels of individual channels - we will refer to it as unknown noise level problem (UNL problem). As expected, in both problems we can approach rate as if both sides would have the missing information.

The article is constructed in the following way. The JRC error correction setting will be formulated in Section \ref{sec2}. There will be introduced the considered encoding scheme and separate decoding procedures in two cases: for undamaged packets only and for the general case (with included error correction). The undamaged case will be discussed in Section \ref{sec3} - both theoretically and comparing with experimental tests. This section begins with straightforward decoding: when we need to consider only a single candidate. Then a general undamaged case is discussed, where it turns out that we need to consider only a small number of candidates per step ($\leq 2$ on average). Section \ref{sec4} contains theoretical analysis and results of experiments for the general damaged case, where the number of candidates per step is approximately described by Pareto distribution with coefficient found using Renyi entropy. Some idealized UNL problem will be analyzed in Section \ref{ucn} to understand performance of different possibilities, especially the concatenation of FC and JRC. Finally, Section \ref{sec5} summarizes the new possibilities of JRC coding paradigm and discusses some further research questions and improvement opportunities.

\section{Joint Reconstruction Codes (JRC)} \label{sec2}
We will now formulate the error correction setting and briefly introduce coding/decoding methods. Details can be found in the accompanied implementation~\cite{impl}. The presented approach is expansion of the method described in~\cite{cortree}, where more discussion can be found, especially for bidirectional correction.
\begin{figure*}[t!]
    \centering
        \includegraphics[width=16cm]{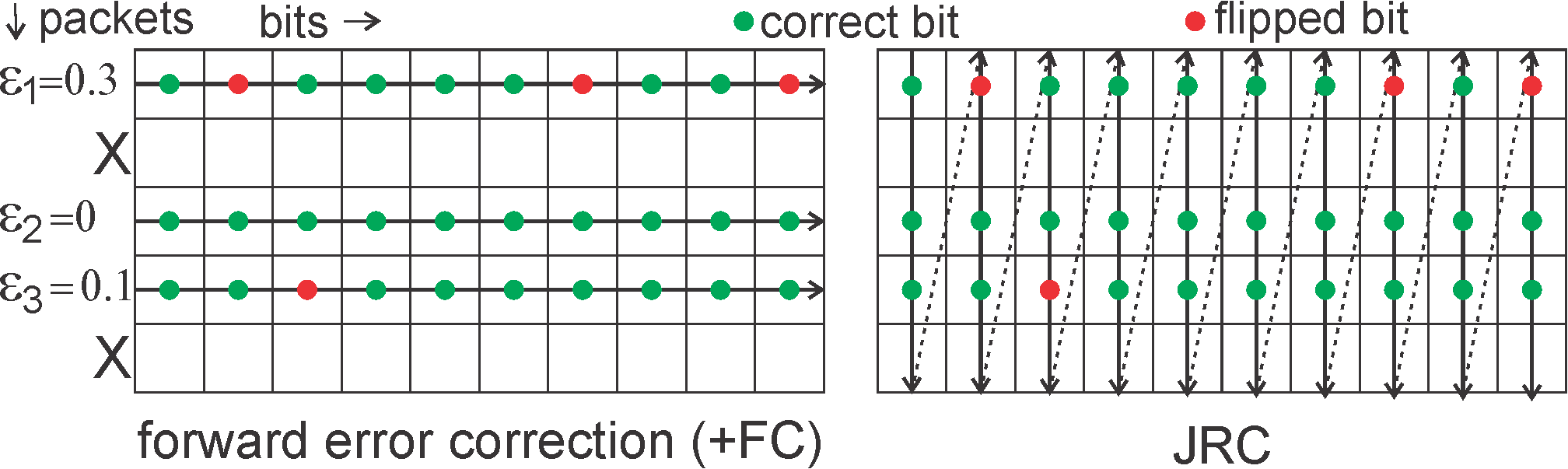}
        \caption{Two approaches for combining the possibilities of fountain codes with error correction (two "X"-marked packets have been lost) for $M=3$ received packets. Left: correct each packet independently, then use FC. Right: JRC tries to reconstruct successive bits of the message from bits on the corresponding position of simultaneously all packets.}
        \label{genpict}
\end{figure*}
\subsection{Encoding}
Let us start with the fountain code (FC) situation. Assume the size of the message is $NL$ and we are receiving some $M$ packets of length $L$ bits from a larger set of packets. The minimal requirement for reconstruction is $M\geq N$. We will now allow the packets to be damaged, as depicted in Fig. \ref{genpict}. For simplicity there will be assumed binary symmetric channel (BSC): every bit of packet $i\in[1,M]$ has $\epsilon_i$ probability of being flipped. If we would apply forward error correction for (a priori) known $\epsilon$ values, the minimal requirement for reconstruction is $\sum_i (1-h(\epsilon_i))\geq N$. We will get the same bound for JRC without the a priori knowledge.

\begin{figure*}[t!]
    \centering
        \includegraphics[width=16cm]{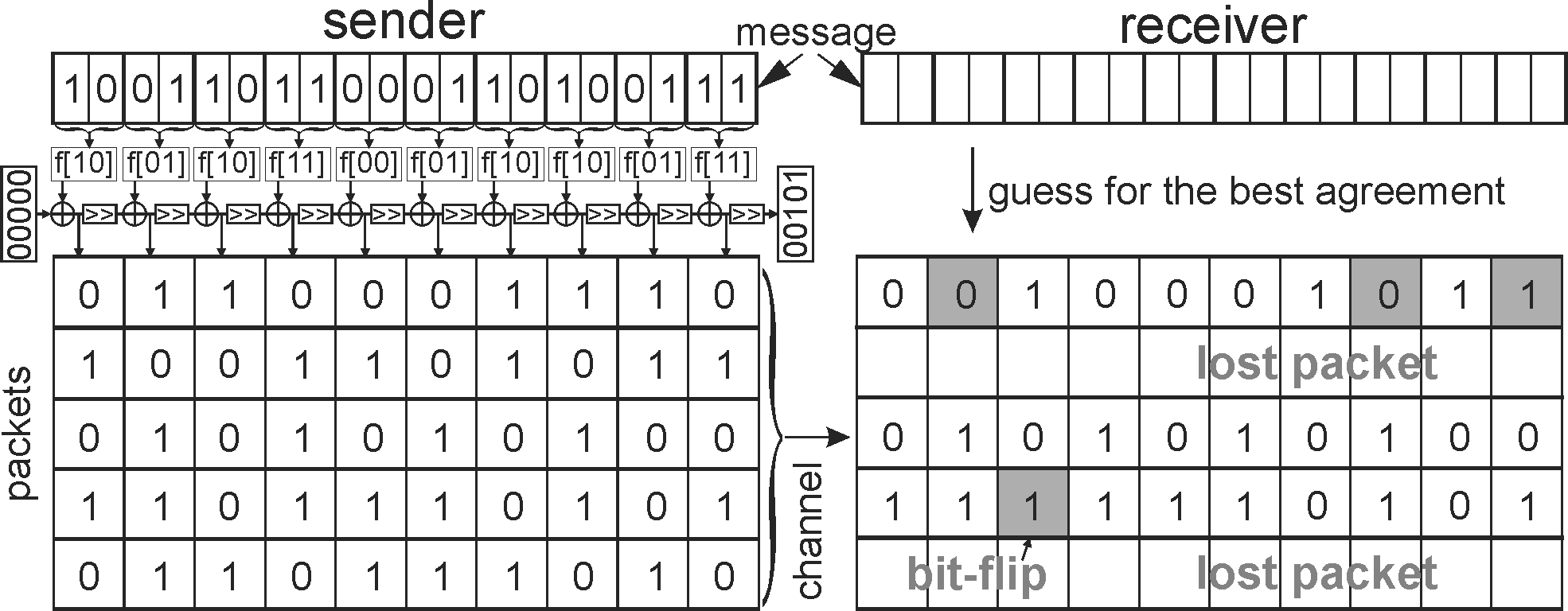}
        \caption{Example of encoding process for 5 bit state and 5 packets, $N=2$ size bit blocks, $M=3$ received packets and $f[00]=01100, f[01]=10010, f[10]=01010, f[11]=10110$ transition function, chosen in a pseudorandom way. The received packets are e.g. $which[1]=1,\ which[2]=3,\ which[3]=4$, the $extract:\ \{0,\ldots, 31\}\to\{0,\ldots,7\}$ function returns 1,3,4 bits of 5 bit state here. Symbol $\oplus$ denotes XOR, "$>>$" denotes cyclic shift right by 1 position of the 5 bit state.}
        \label{example}
\end{figure*}

Let us divide the message into length $N$ bit blocks, each block corresponds to a single bit in every packet (vertical lines in Fig. \ref{genpict}). Specifically, the encoding procedure has $L$ steps. In $k$-th encoding step ($k\in\{0,\ldots,L-1\}$) there are used $\{Nk, ..., N(k+1)-1\}$ bits of the message (bit block $x$), to produce $k$-th bit of every packets. There is required an internal state of encoder to connect redundancy of successive blocks. The current implementation uses 64 bit state for this purpose, producing $k$-th bit of $i$-th packet as $i$-th bit of the $state$. This way the number of produced packets is limited to 64, the actually received packets correspond to some subset of these 64 bits. The internal $state$ needs to be modified accordingly to the currently encoded $N$ bit block $x$. There is used a pseudorandomly chosen transition function $f:\{0,\ldots, 2^N-1\}\to\{0,\ldots,2^{64}-1\}$ for this purpose. The $state$ transition is a cyclic shift of ($state$ XOR $f[x]$). The encoding procedure is schematically presented as Method \ref{enc}, example of its application is presented in Fig. \ref{example}. Observe that using a pseudorandom number generator (PRNG) initialized with a cryptographic key to choose the $f$ function, we could include encryption in such encoding.

Finally encoding procedure is: set $state$ as some arbitrary $initial\_state$ (known to receiver), then perform encoding step for blocks $0$ to $L-1$. The final state: $final\_state$ would be beneficial for decoder for the final verification of unidirectional decoding, and is necessary for bidirectional decoding. It may be included in the header of packet. However, unidirectional correction can be performed without it, at cost of probable damage of some last bits of the message.

\begin{algorithm}[htbp]
\footnotesize{
\caption{Encoding procedure}
\label{enc}
\begin{algorithmic}
\REQUIRE $state=initial\_state$
\FOR {$k=0$ to $L-1$}
\STATE set $x$ as $k$-th $N$ bit block to encode: \\ \qquad $x=\{Nk, ..., N(k+1)-1\}$ bits of the message
\STATE $state = state$  XOR $ f[x]$  \qquad\qquad \COMMENT{state transition}
\STATE for all $i$, set $k$-th bit of packet $i$ as $i$-th bit of $state$:\\
\qquad \qquad$packet[i]_k = state_i$   \qquad \COMMENT{$x_i$ is $i$-th bit of $x$}
\STATE $state = state >> 1$ \qquad \COMMENT{cyclic shift by one position}
\ENDFOR
\STATE $final\_state=state$        \qquad \COMMENT{it can be useful for decoding}
\end{algorithmic}
}
\end{algorithm}

\subsection{Decoding for undamaged case - no error correction}
The decoding procedure searches for a message leading to all the received packets by expanding a tree of promising candidates. Depth $k$ node of this tree represents a candidate of length $Nk$ prefix of the message, corresponding to $k$ length prefixes of all the received packets. Expanding a node denotes creating its children corresponding to $N$ bit block longer prefix of the message.

In the undamaged case we consider only candidates having complete agreement with all the received packets, the remaining expansions are not considered. In this case, a natural approach is to consider a list of all candidates up to a given position and successively shift this position. The candidate list for the next position is obtained by expanding all candidates from the previous list by a single step, considering only those being in agreement with the corresponding bit of all received packets.

For performance reasons, this method should have prepared a list of allowed single step expansions for all cases. We have received some $M$ of e.g. 64 packets. Denote $which[i]\in\{1,\ldots,64\}$ for $i=1,\ldots,M$ as the original number of $i$-th received packet (its bit position). Denote by $data[k]$ the $M$ bit sequence of $k$-th bit of all received packets:
$$data[k]_i = received\_packet[i]_k$$
for $i=1,\ldots,m,\ k=0,\ldots,L-1$, where $x_i$ denotes $i$-th bit of $x$ here. Denote by $extract(state)$ as $M$ bit sequence of bits of $state$ on positions corresponding to all received packets:
$$extract(state)_i=state_{which[i]}\qquad \textrm{for}\ i=1,\ldots,M$$
Now a node can be expanded by sequence $x$ if $data[k]=extract$($f[x]$ XOR $state$), or in other words if
$$extract(f[x])= extract(state)\  \textrm{XOR}\ data[k]\quad :=z$$
Hence, the current situation can be described by $z$ which will be referred as a syndrome.
We can have prepared $candidates[z]$ table containing a list of $x$ being in agreement with syndrome $z\in\{0,\ldots, 2^M-1\}$. The decoding method is schematically presented as Method \ref{dec1}. Its performance is discussed in Section \ref{sec3}.

\begin{algorithm}[htbp]
\footnotesize{
\caption{Decoding for undamaged case}
\label{dec1}
\begin{algorithmic}
\REQUIRE $current\_list$ = $\{initial\_state\}$ \qquad \COMMENT{root of the tree}
\REQUIRE $candidates[z]$: list of all $x$ such that $extract(f[x])=z$
\FOR{$k=0$ \TO $L-1$ }
\STATE empty $new\_list$
\FOR {every $node$ in $current\_list$ }
\STATE $cnd$ = $candidates[extract(node.state)$ XOR $data[k]]$
\STATE append $cnd$ with updated $state$s to $new\_list$ 
\ENDFOR
\STATE $current\_list$ = $new\_list$
\ENDFOR
\STATE trace back any coding from the final $new\_list$ (with $state=final\_state$ if available)
\end{algorithmic}
}
\end{algorithm}

For each of $2^M$ syndromes $z$, there are possible $2^N$ candidates, each of them has $2^{-M}$ probability of agreeing with all the received packets. Hence, the expected size of the list pointed by $candidates[z]$ is $2^N/2^M$. Finally, the total size of all lists of candidates and their pointers is $O(2^N+2^M)$ here. This exponential growth restricts to reconstruction only from a relatively small $M$. It could be slightly relaxed for example by splitting the $M$ packets into a few subsets and use a separate $candidates$ table for each of them, intersecting their lists.\\

\textbf{Interleaving}. A more efficient way for increasing the number of packets for reconstruction is splitting the decoding step into $S\in\mathbb{N}$ substeps (phases), which use disjoined subsets of packets. These substeps should be performed cyclically, hence we will refer to this approach as interleaving. The phase of encoding/decoding is indexed by $0\leq s < S$ number of substep, and $s$ increases cyclically modulo $S$ in successive steps.

We would like that such $S$ successive substeps correspond to a single original step - all further analysis should be seen that, in the case of interleaving, a single step is in fact grouped $S$ successive substeps. Let us split the original $N$ bit blocks of the message per step, into bit blocks used in separate phases: $N=\sum_s N_s$ ($N_s\in\mathbb{N}$), which can be equal. The encoding Method \ref{enc} corresponds to a single substep now, evolving the same $state$ in succeeding phases. In phase $s$ it uses successive $N_s$ bit block of the message to produce successive bit of $s$-th subset of packets. Hence, in the current setting, it could produce $64S$ packets. If all $N_s$ are equal, we can use the same transition function $f:\{0,\ldots,2^{N_s}\}\to \{0,\ldots,2^{64}-1\}$ for all phases.

The decoding analogously goes cyclically through these $S$ phases, considering all possible candidates up to position corresponding to a given substep. The received packets are now from $S$ subsets: $M=\sum_s M_s$, where $M_s$ is the number of received packets from $s$-th subset. Now we need two-dimensional $candidates[s][z]$ table, containing all possible candidates for the current phase ($s$) and syndrome ($z$). Analogously to the non-interleaving case, we can estimate the memory requirement for all lists and pointers as $O\left(\sum_{s=0}^S 2^{N_s}+2^{M_s}\right)$. This formula is minimized for uniform split of both $N$ and $M$. The former can be obtained by choosing all $N_s$ equal. Uncontrolled nature of received packets might be an issue. We could enforce upper bound for $M_s$ by limiting the number of packets from a single phase, or eventually use a few $candidates$ tables for a given phase and intersect their lists.

\subsection{Decoding in the general case (with error correction)}
In the general case we allow for damaged packets: tolerate some disagreement with the received packets while reconstruction. The higher assumed (a posteriori) damage level ($\epsilon_i$), the higher tolerance for disagreement of bits of a given packet. Considering all candidates up to a given position, like in Method \ref{dec1}, would be completely impractical here as their number would grow exponentially. Instead, in every step we will find and expand only the most promising node, as the one maximizing weight $W$, which is logarithm of the probability of being the proper node, obtained using Bayesian analysis. As discussed in \cite{cortree}, this probability is a product of probability of the assumed error vector and one over probability of accidental agreement of incorrect nodes.

Assuming that a given node is incorrect, the probability of accidental agreement of some of its $2^N$ possible expansion with $M$ bits of received packets is $2^N/2^M$. Let us denote $k$-th step error vector for a given message candidate by $\mathbf{E}^k=\{E_i^k\}$. $E_i^k=0$ if $k$-th bit of $i$-th packet agrees, 1 otherwise:
$$\mathbf{E}^k = extract(state \ \textrm{XOR}\ f[x])\ \textrm{XOR}\  data[k]$$
Now assuming that $i$-th packet came through $\epsilon_i$ BSC, probability of $\mathbf{E}^k$ error vector is:
\be \Pr(\mathbf{E}^k)=\prod_{i=1}^M P_{\epsilon}(E_i^k) \ee
 $$\textrm{where}\quad P_{\epsilon}(0):=1-\epsilon\quad,\quad P_{\epsilon}(1):=\epsilon$$
We can finally write the formula for weight we want to maximize:
$$ W=\sum_k W(\mathbf{E}^k)$$ 
\be W(\mathbf{E}^k) = M - N + \sum_{i=1}^M \lg(P_{\epsilon_i}(E_i^k))\ee
In a given moment, the most probable node to expand is the one having the largest weight $W$. The $M-N$ term favors longer expansions, the error vector term contains a penalty for disagreement with the received damaged packets. The weight of expansion of a node is the weight of its parent plus $W(\mathbf{E}^k)$ of the currently considered expansion, evaluating disagreement with $k$-th bit of the packets.

In practical applications we need to quickly find the best candidates for given expansion. As for the undamaged case, the current situation is sufficiently described by syndrome $z=extract(state)$ XOR $data[k]$. This time $candidates[z]$ lookup table should contain $2^N$ bit block candidates $x$, sorted by the $W(z$ XOR $extract(f[x]))$ weight. The final decoding procedure is schematically presented as Method \ref{dec2}, detailed implementation is in \cite{impl}, its performance is discussed in Section \ref{sec4}.

\begin{algorithm}[htbp]
\footnotesize{
\caption{Sequential decoding for the general case}
\label{dec2}
\begin{algorithmic}
\REQUIRE empty $heap$ of nodes to consider
\REQUIRE $candidates[z]$: list of $2^N$ blocks $x$ \\ \qquad\qquad\qquad\qquad\qquad sorted by $W(z$ XOR $extract(f[x]))$
\STATE insert root to $heap$: $initial\_state$, $k=0$ position, $W=0$ weight
\REPEAT
\STATE $node$ = getMax($heap$)\quad \COMMENT{retrieve max weight node from $heap$}
\STATE if applicable, add to $heap$ \\ \qquad\qquad\qquad the next of $candadates$ of parent of $node$
\STATE add first of $candidates$ of $node$ to $heap$
\UNTIL{reached final position (and if available: $state=final\_state$)}
\end{algorithmic}
}
\end{algorithm}

This time $candidates$ table has much larger memory requirements: $O(N\cdot 2^{M+N})$ as we need all sorted $x\in\{0,\ldots,2^N-1\}$ for every syndrome $z\in\{0,\ldots,2^M-1\}$. There could be used approximated scheme to slightly relax the restriction for the number of packets for reconstruction. For example, to approximate the optimal order, one could use $candidates$ table only for some number of the least damaged of received packets (as they have the highest impact on the weight). Then all the received packets can be used to calculate the weight of a given extension.

Using interleaving, the total memory cost is analogously $O\left(\sum_s N_s\cdot 2^{M_s+N_s}\right)$. As previously, $N_s$ should be chosen equal, while random nonuniform split of $M$ might be an issue, limiting the maximal number of received packets from a single phase used for reconstruction. As in the non-interleaved case, if exceeding the assumed resources, the least damaged should be used to approximate the optimal order.
\begin{table*}[t!]
\caption{Probability of straighforward decoding for random transition function (eq. (\ref{prsd})).}
\label{strdec}
\centering
\begin{tabular}{|c|c|c|c|c|c|c|c|c|}
  \hline
    $N$ & $M=1$ & $M=2$ & $M=3$ & $M=4$ & $M=5$ & $M=6$ & $M=7$ & $M=8$ \\\hline
  1 & 0.5 & 0.75 & 0.875 & 0.9375 & 0.96875 & 0.98436 & 0.99219 & 0.99609 \\
  2 & - & 0.09375 & 0.41016 & 0.66650 & 0.823059 & 0.0.90891 & 0.953794 & 0.97673 \\
  3 & - & - & 0.00240 & 0.12082 & 0.38572 & 0.634028 & 0.79999 & 0.89542 \\
  4 & - & - & - & $1.1\cdot 10^{-6}$ & 0.01040 & 0.12901 & 0.37613 & 0.61971 \\
  5 & - & - & - & - & $1.8\cdot 10^{-13}$ & $7.6\cdot 10^{-5}$ & 0.01442 & 0.13236 \\
  6 & - & - & - & - & - & $3\cdot 10^{-27}$ & $4.2\cdot 10^{-9}$ & 0.00018 \\
  7 & - & - & - & - & - & - & $7\cdot 10^{-55}$ & $1\cdot 10^{-17}$ \\
  8 & - & - & - & - & - & - & - & $3\cdot 10^{-110}$ \\
  \hline
\end{tabular}
\end{table*}

\section{Undamaged packets case}  \label{sec3}
Let us now analyze the situation with only undamaged packets - analogous to fountain codes, no error correction included. We can use simpler decoding Method \ref{dec1} in this case: consider all candidates up to a given position and successively shift this position.

\subsection{Straightforward decoding}
Let us look at $\{extract(f[x])\}_{x=0,\ldots,2^N-1}$ length $M$ sequences: containing bits of $f[x]$ corresponding to all possible bit blocks of the message ($x$). Observe that if all these sequences turn out different, the block of $k$-th bit of the received packets ($data[k]$) immediately determines the unique $x$ to decode: $candidates[extract(state)$ XOR $data[k]]$ is a single value. We will refer to this case as \emph{straightforward decoding} case: we can directly determine the only possible candidate, move to successive position and so on as in Method \ref{dec3}. The number of decoding steps is exactly the number of encoding steps in this case.

\begin{algorithm}[htbp]
\footnotesize{
\caption{Decoding for straighforward case}
\label{dec3}
\begin{algorithmic}
\REQUIRE $state=initial\_state$
\REQUIRE $candidates[z]$: the only $x$ such that $extract(f[x])=z$
\FOR{$k=0$ \TO $L-1$ }
\STATE $x=candidates[extract(node.state)$ XOR $data[k]]$
\STATE produce $x$\qquad\qquad \qquad\qquad\qquad \qquad \COMMENT{decoded symbol}
\STATE $state= (state\ \textrm{XOR}\ f[x]) << 1$\qquad\qquad \COMMENT{cyclic shift}
\ENDFOR
\end{algorithmic}
}
\end{algorithm}

Hence, assuming a completely random transition function $f$ ($\Pr(0)=\Pr(1)=1/2$), we can find probability of such straightforward decoding: that $2^N$ random length $M\geq N$ sequences will be different. Table \ref{strdec} contains some its numerical values.
\be \Pr(\textrm{straighforward decoding})=\frac{\left(2^M\right)!/\left(2^M-2^N\right)!}{\left(2^M\right)^{(2^N)}} \label{prsd}\ee
Assuming a completely randomly chosen transition function $f$, for some parameters like $M>>N$ we can get straightforward decoding with a reasonable probability. \\

In many scenarios we could improve this probability by designing transition function and influencing the subset of packets used for reconstruction. Let us discuss such possibility for the following practical example:

\textbf{Storage media protection problem}: we would like to store a message on a few storage media like DVD disks, such that we can retrieve the message independently from any single one of them as long as there is no damage, or it is low. Additionally, for high damage, we should be able to retrieve the message from a few of these storage media.

For speed purposes, let say we would like to straighforward decode $N=8$ bits/step from such single undamaged storage medium. It can be done by storing 8 JRC packets on such single medium, and choose the transition function as a (random) permutation of bits corresponding to these packets. In other words, $extract[f(x)]:\{0,\ldots, 255\}\to \{0,\ldots, 255\}$ should be chosen as a permutation for $extract$ choosing bits from a single medium. This way we get straighforward decoding from such 8 undamaged packets. If we want to be able to repair some damage for such single medium, we can place more of these packets there.

For example imagine recording 4 disks, using 16 (of 64) different packets for each of them. Now for the 8 undamaged packets of any of these disks we can straightforward decode the message. Using all 16 packets from one medium, we get rate 1/2 forward error correction (maximum bit-flip probability $\epsilon\approx 0.110$). Using 64 packets from all 4 disks, we get rate 1/8 forward error correction (maximum bit-flip probability $\epsilon\approx 0.295$). Hence, if encoding a message this way, JRC allows here for 3 levels of decoding depending on available resources.
\subsection{General undamaged case}
Let us define \emph{width}, characterizing the decoding process as the average number of candidates considered per position

$$width = \frac{\textrm{number of decoding steps}}{\textrm{number of encoding steps}}=$$
$$=\frac{\#\textrm{ nodes considered while reconstruction}}{L}$$

The number of encoding steps for $LN$ bit message is $L$. The number of decoding steps is the number of considered candidates: nodes of reconstruction tree.

For straightforward decoding $width=1$, which is the lower bound. We are interested in understanding the probability distribution of $width$ for given reconstruction parameters. As it will be explained, we should expect approximately Gaussian distribution for the undamaged $M>N$ case, what is confirmed by tests. For damaged case: with included error correction, we get approximately Pareto distribution of $width$, what will be discussed in Section \ref{sec4}.

\begin{figure*}[t!]
    \centering
        \includegraphics[width=16cm]{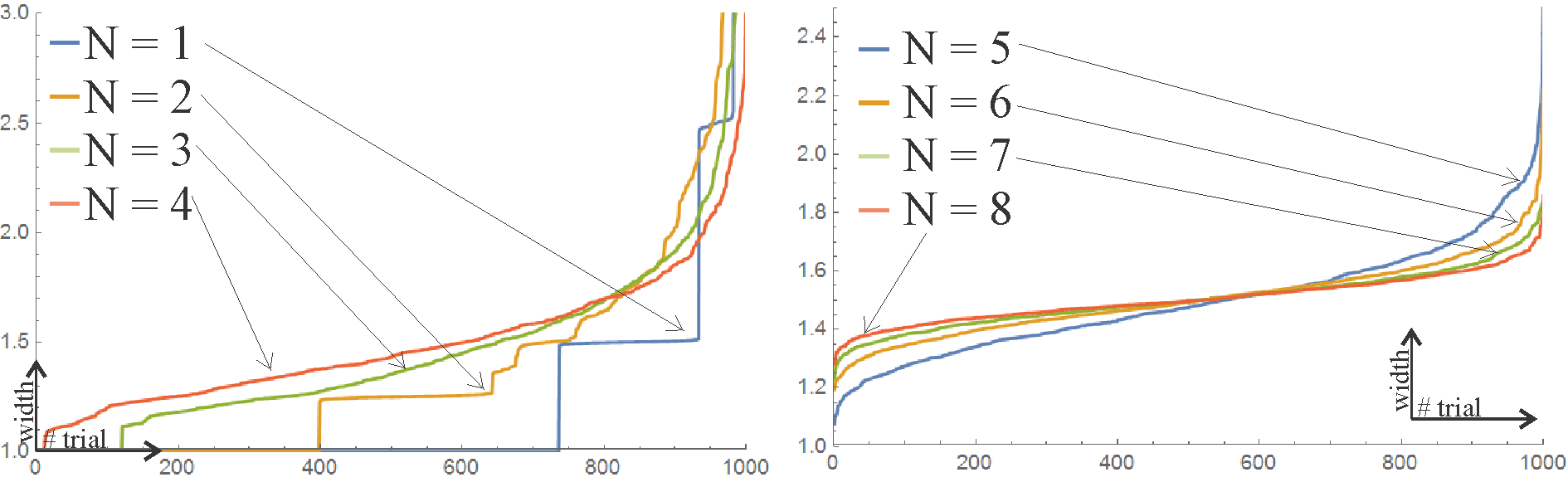}
        \caption{Experimental results for Method \ref{dec2}, $N=1,2,3,4,5,6,7,8$ required packets and received $M=N+1$ undamaged packets. The graphs show sorted $widths$ for 1000 trials for length 1000B messages ($L=\lceil 8000/N\rceil$ steps).}
        \label{undam}
\end{figure*}

Figure \ref{undam} shows $width$s for $M=N+1$ cases and $N=1,\ldots ,8$. It was obtained by performing 1000 reconstructions for random conditions (message and subset of packets) using Method \ref{dec2}, and then sorting obtained $width$s. If we would switch both axes of this plot and normalize, we would get empirical approximation of the cumulative distribution function for $width$. In its left panel (low $N$) we can see some $width=1$ lines corresponding to straightforward decoding - up to approximately 0.75, 0.41, 0.12, 0.01 of 1000 for correspondingly $N=1,2,3,4$, as we can read from Table \ref{strdec}.

For a larger $N$ we can see a shape resembling cumulative distribution function of Gaussian distribution. Let us find this behavior assuming some randomness. Imagine that in a given step we have considered $i$ candidates. They have $2^N i$ possible expansions - one of them is the proper message, the remaining have $2^{-M}$ probability of accidentally agreeing with $M$ bits from the received packets. Finally, the probability that there will be $j$ candidates to consider in the next position is
\be p_{ij}={2^N i -1 \choose j-1} 2^{-M(j-1)} \left(1-2^{-M}\right)^{2^N i-j } \ee
This stochastic matrix (of infinite size) allows to find the stationary probability distribution for the number of candidates to consider: such that $p_i =\sum_j p_j p_{ij}$, where $p_i$ is the probability that there will be $i$ candidates in a given position. This $\{p_i\}_{i=1}^\infty$ distribution tends to 0 for $M > N$ (not true for $M=N$). In the $M=N+1$ case it stabilizes for large $N$ on approximate distribution:
$$\{0.41884,0.32221,0.15680,0.06446,0.02438,$$ $$0.00875,0.00303,0.00102,0.00034,0.00011,\ldots\}$$
with expected value $2$ and standard deviation $3.28$. For $M=N+2$ case the asymptotic distribution is approximately:
$$\{0.72383,0.22726,0.04175,0.00620, $$ $$0.00084,0.00011,0.00001,\ldots \}   $$
with average value $1.33$ and standard deviation $2.58$.

Finally, for the $M=N+1$ case and $N>\approx 4$, the width of Method \ref{dec1} should be approximately a Gaussian distribution with expected value 2. It means that per position there is 1 proper candidate and on average 1 improper. However, Figure. \ref{undam} clearly shows expected value 1.5 - the difference is caused by using Method \ref{dec2} for these results. This algorithm considers candidates of various lengths, favoring the longest ones in the undamaged case. Hence, after considering the proper candidate, it will not consider the remaining candidates on a given position here. The proper candidate is on average in the middle, hence the expected width for Method \ref{dec2} should be 1 (proper) + 0.5 (improper) = 1.5 here, what agrees with the Figure. \ref{undam}.

One could wrongly conclude that it is better to use Method \ref{dec2} for the undamaged case to reduce the linear coefficient e.g. from 2 to 1.5 here. However, this algorithm requires finding the best node for each step, making it more costly than Method \ref{dec1}.

\subsection{Conclusions and comparison with fountain codes}
In this section we have discussed proposed coding approach as a direct replacement for fountain codes: for undamaged packets. For some probability, $M=N$ packets are sufficient to fully reconstruct the message, especially for small $N$. For the $M=N+1$ case: one excess packet, the decoding is very cheap: requires on average twice more steps than encoding. Probability of failure is practically zero.

In contrast, FC have relatively large probability of failure for a small number of excess packets (the used random matrix is not invertible), like $\approx 50\%$ for the $M=N+1$ case~\cite{mackay}. However, they can inexpensively operate on arbitrarily large number of packets, what is costly for JRC approach.

Another advantage of FC is retrieving a part of the message if complete reconstruction is not possible. This subset is usually random, however, more important packets can be better protected in unequal error protection~\cite{unequal} variant of FC. In contrast, JRC tries to reconstruct the entire message. If it fails: exceeds the assumed resources, the beginning of the message up to this position (and the end in bidirectional correction), are nearly properly reconstructed.

\section{General damaged case} \label{sec4}
We will now discuss the performance of the general case: with allowed damaged packets. This time decoding is much more costly due to included error correction. While for undamaged case the $width$ had approximately Gaussian distribution with a small expected value ($\approx 2$ for the $M=N+1$ case), the $width$ for damaged case is approximately described by Pareto distribution with parameter $c\geq 0$: increasing twice the $width$ limit decreases probability of failure approximately $2^c$ times. The $c=0$ case is the Shannon theoretical rate limit. The $c=1$ case is called cutoff rate: the maximal rate with finite asymptotic $width$ ($\int_0^\infty x^{-c}dx=\infty$ for $c\leq 1$). We will now find $c$ for a general case. Derivation for simpler case: single BSC and test results can be found in \cite{cortree}.
\begin{figure*}[t!]
    \centering
        \includegraphics[width=16cm]{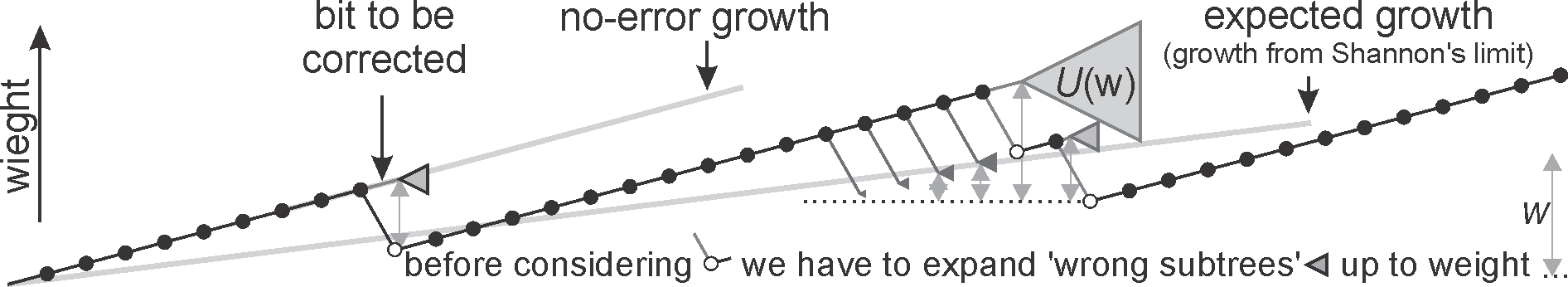}
        \caption{Schematic picture of the proper reconstruction we would like to find. After an error, we expand improper candidates until reaching the weight of the proper one. We need to find the expected size $(U(w))$ and probability distribution $(V(w))$ for these subtrees of improper candidates.}
        \label{an}
\end{figure*}

\subsection{Finding the Pareto coefficient $c$}
Let us remind weights found in Section \ref{sec2}: quantitatively describing how likely we will tolerate bit-flips depending on damage level of each packets (that $i$-th packet came through $\epsilon_i$ BSC). Denote error vector as $\mathbf{E}^k=\{E_i^k\}_{i=1..M}$, where $E^k_i=0$ if $k$-th bit of $i$-th packet agrees, $E^k_i=1$ otherwise. Now the weight of $k$-th block of candidate is $W(\mathbf{E}^k)$:
$$W(\mathbf{E}^k)=M-N + \sum_{i=1}^M \lg(P_{\epsilon}(E_i))$$
$$ \textrm{where}\ P_{\epsilon}(0):=1-\epsilon,\ P_{\epsilon}(1):=\epsilon.$$
The total weight of a candidate is the sum of its $W(\mathbf{E}^k)$. In every step of Method \ref{dec2} we choose the looking most probable node of the tree (candidate): having the largest weight $W$, and try to expand this node.

The Shannon rate limit is $\sum_{i=1}^M (1-h(\epsilon_i))\geq N$, what is equivalent to $\sum_{\mathbf{E}} \Pr(\mathbf{E})W(\mathbf{E})\geq 0$.
In other words, for rates below the Shannon limit, the weight of the proper candidates is on average growing. In contrast, the weights of improper candidates is on average decreasing, allowing to cut these branches of reconstruction tree.

\begin{figure*}[!t]
\centering
\begin{IEEEeqnarray}{rCl}
U(w):=\textrm{the expected number of improper nodes in subtree with }w\textrm{ weight drop} \nonumber\\
U(w)=\left\{\begin{array}{ll} 1+2^N \sum_{\mathbf{E}\in\{0,1\}^M} U(w+W(\mathbf{E}))/2^M \quad & \textrm{for }w\geq 0\\
                 0 & \textrm{for }w<0 \end{array}\right. \label{vequ}
\end{IEEEeqnarray}
\begin{IEEEeqnarray}{rCl}
V(w):=\Pr(\textrm{weight on the correct path will drop by at most }w) \nonumber \\
V(w)=\left\{\begin{array}{ll} \sum_{\mathbf{E}\in\{0,1\}^M} \left(\prod_{i=1}^M P_{\epsilon_i}(E_i)\right)\cdot V(w+W(\mathbf{E}))\quad & \textrm{for }w\geq 0\\
                 0 & \textrm{for }w<0 \end{array}\right. \label{veqv}
\end{IEEEeqnarray}
\hrulefill
\end{figure*}

The situation is depicted in Fig. \ref{an}. The dots correspond to length $N$ bit blocks of the message. The black like represents the (unknown) proper reconstruction we are looking for. As long as there are no errors, the weight is growing ($W(\mathbf{E}_k)>0$), hence it will remain the favorite candidate of sequential decoding. However, the weight drops while errors appear ($W(\mathbf{E}_k)<0$). In this case, the decoder will try different (improper) candidates, until exploring all possibilities having higher weight than the weight of the proper candidate - "wrong subtrees": triangles in the figure. Hence, what is crucial is the weight drop $w$ parameter (arrows in the figure): the maximal weight on the proper path up to the considered position minus the current weight. Without error, the currently considered node has the highest weight so far: $w=0$. However, larger $w$ can theoretically also appear, with probability decreasing exponentially with $w$. Situation with $w$ parameter requires considering subtrees of improper candidates of size growing exponentially with $w$.\\

We will now find these two behaviors. Let us start with finding the expect number of improper nodes to expand for a given weight drop. Define $U(w)$ as the expected number of nodes in subtree with $w$ weight drop.
If a given node is created, we potentially need to expand all of its $2^N$ children.
Let us assume that each of them corresponds to a completely random $\mathbf{E}=\{E_i\}_i$ vector: $\Pr(\mathbf{E})=1/2^M$.
Finally, the expected number of nodes to expand for each of $2^N$ possibilities is
$\sum_{\mathbf{E}\in\{0,1\}^M} U(w+W(\mathbf{E}))/2^M$, getting functional equation (\ref{vequ}).
Large $w$ behavior is crucial here - it corresponds to rare large subtrees of improper candidates. In this case we can neglect the "$1+$" term and the remaining linear functional equation should have asymptotic behavior of form:
$$ U(w)\propto 2^{uw}\qquad \textrm{for }w\to\infty \quad \textrm{and some}\quad u>0$$
Inserting this assumption to (\ref{vequ}) we get:
$$2^{uw}=2^{N-M} \sum_{\mathbf{E}\in\{0,1\}^M} 2^{u(w+M-N+\sum_i \lg(P_{\epsilon_i}(E_i)))}$$
$$2^{(M-N)(1-u)}=\sum_{\mathbf{E}\in\{0,1\}^M} \prod_i P_{\epsilon_i}(E_i)^u=$$
$$=\prod_{i=1}^M \left(\epsilon_i^u+(1-\epsilon_i)^u\right) $$
$$ N=M-\sum_{1=1}^M \frac{\lg(\epsilon_i^u+(1-\epsilon_i)^u)}{1-u}$$
\be N=\sum_{i=1}^M 1-h_u(\epsilon_i)\label{equ}\ee
where $h_u(\epsilon) = \frac{\lg(\epsilon^u+(1-\epsilon)^u)}{1-u}$ is the Renyi entropy.\\

Let us now find the probability distribution of values of $w$ obtained while reconstruction. Define $V(w)$ as CDF for $w$: the probability that weight on the correct path will drop by at most $w$.
Comparing situation in successive positions and using BSC assumption: $\Pr(\mathbf{E})=\prod_{i=1}^M P_{\epsilon_i}(E_i)$, we get (\ref{veqv}) functional equation.
For $w<0$, $V(w)=0$. As discussed, $0<V(0)$ is the probability that a proper node will have the highest weight so far. From definition, $\lim_{w\to\infty} V(w)=1$. Finally, we can assume that the functional equation has solution of form:
$$ 1-V(w)\propto 2^{-vw}\qquad \textrm{for }w\to\infty \quad \textrm{and some}\quad v>0$$
Inserting this assumption to (\ref{veqv}) we get:
$$2^{-vw} = \sum_{\mathbf{E}} \left(\prod_i P_{\epsilon_i}(E_i)\right) 2^{-v(w+M-N+\sum_i \lg(P_{\epsilon_i}(E_i)))}$$
$$2^{v(M-N)}=\sum_{\mathbf{E}\in\{0,1\}^M} \prod_i P_{\epsilon_i}(E_i)^{1-v}=$$
$$= \prod_{i=1}^M \left(\epsilon_i^{1-v}+(1-\epsilon_i)^{1-v}    \right)$$
$$ N=M-\sum_{1=1}^M \frac{\lg(\epsilon_i^{1-v}+(1-\epsilon_i)^{1-v})}{v}$$
\be N=\sum_{i=1}^M 1-h_{1-v}(\epsilon_i)\label{eqv}\ee
comparing with formula (\ref{equ}) for $u$, we see that $v=1-u$.\\

Having the exponent coefficients of $U(w)$ and $V(w)$ functions from (\ref{equ}) and (\ref{eqv}), we can assume that $U(w)\approx  c_u 2^{uw},\ V(w)\approx 1-c_v 2^{-vw}$ for some unknown coefficients $c_u,\ c_v$. Now the asymptotic probability that the number of nodes of subtree of wrong correction for a given position will exceed some number of steps $s$ is:
$$\Pr(\#\textrm{ nodes}>s)\approx \Pr\left(w>\frac{\lg(s/c_u)}{u}\right)\approx $$
$$\approx 1-c_v 2^{-\frac{v}{u}\lg(s/c_u)}= 1-c_v \left(\frac{s}{c_u}\right)^{-v/u}$$
\be\Pr(\#\textrm{ nodes in improper subtree}>s)\approx 1-c_p s^{-c}\ee
where $c:=v/u=1/u-1$ (i.e. $u=1/(1+c)$) and $c_p=c_v/c_u^c$. We have obtained the Pareto probability distribution with exponent which can be found analytically: it is $c$ if
$$N=\sum_{i=1}^M 1-h_{1/(1+c)}(\epsilon_i)$$
In other words, the sum of $R_c(\epsilon)=1-h_{1/(1+c)}(\epsilon)$ over all received bits should be the number of bits of the message.\\
\begin{figure*}[t!]
    \centering
        \includegraphics[width=16cm]{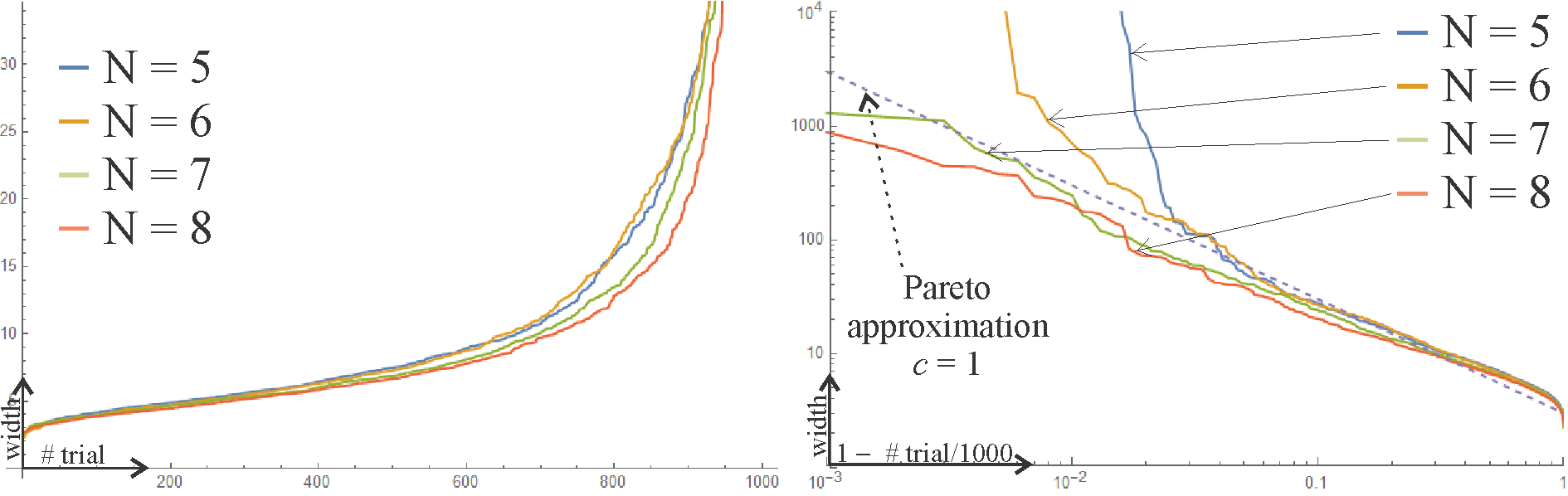}
        \caption{Experimental results (solid lines) and Pareto approximation (dashed line) for $N=5,6,7,8$ required packets, $N-1$ received undamaged and two packets having damage level chosen to obtain $c\approx 1$ Pareto coefficient ($\epsilon=0.05$ and $0.04$ bit-flip probability). While they have the same $c$, larger $N$ allows to stabilize the correction process. Left: sorted average $width$ from 1000 trials. Right: the same plot in log-log scale for number of trial rescaled to $[0,1]$ range.}
        \label{dam}
\end{figure*}
\begin{figure*}[t!]
    \centering
        \includegraphics[width=16cm]{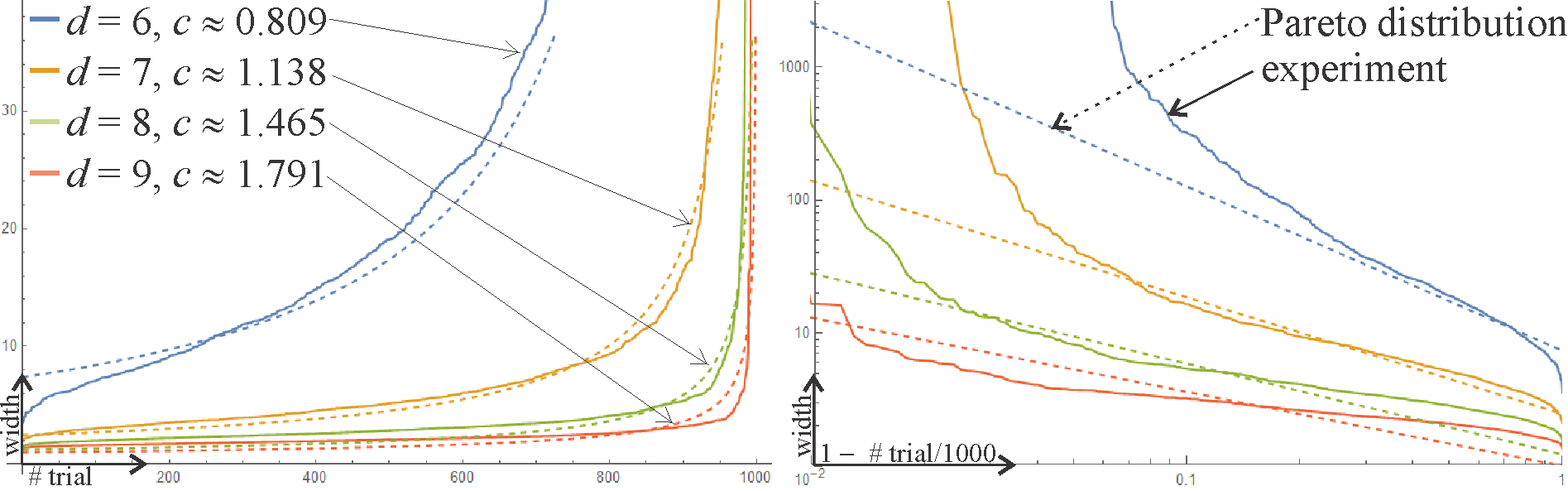}
        \caption{Experimental results (solid lines) and Pareto approximation (dashed lines) for $N=3$ required packets with 2 undamaged and $d=6,7,8,9$ badly damaged packets: with $\epsilon=0.2$ bit-flip probability. Left: sorted average $width$ from 1000 trials. Right: the same plot in log-log scale for number of trial rescaled to $[0,1]$ range.}
        \label{bad}
\end{figure*}

These considerations were for unidirectional decoding: in direction of encoding, starting from the $initial\_state$. As discussed and tested in \cite{cortree} for single BSC, we could also perform bidirectional correction: simultaneously perform backward direction sequential decoding from the $final\_state$. Failure of decoding is caused by critical error concentration (CEC): rare event requiring to consider very large number of improper candidates (large weight drop $w$) - exceeding assumed resource limit. While single CEC prevents successful reconstruction for unidirectional decoding, bidirectional decoding can perform nearly complete reconstruction in this case (and mark the uncertain part). Two CECs are required to essentially cripple the process in this case: probability of such event is approximately square of probability of a single CEC. Finally, Pareto coefficient $c$ practically doubles in this case: $c=1/2$ bidirectional correction has similar behavior as $c=1$ unidirectional.
\subsection{Experimental tests}
Extensive tests for single BSC, including bidirectional correction, can be found in \cite{cortree}. We will now discuss results of two experiments for the general JRC case relying on using multiple packets. Implementation \cite{impl} were used for the tests (unidirectional), both tests use 1000B messages.

The first experiment, presented in Figure \ref{dam}, uses $N-1$ undamaged packets and two damaged packets: of $\epsilon=0.04$ and $0.05$, chosen to get $c\approx 1$ (cutoff rate, $\sum_i R_1(\epsilon_i)\approx N$). Increasing $N$ does not affect $c$, but the graphs show some dependence on $N$: smaller $N$ case has a bad behaving tail, stabilized thanks to using more nodes.

The second experiment, presented in Figure \ref{bad}, tests the exploitation of tiny informational content of highly damaged packets: of $\epsilon=0.2$, there are $d$ of such packets. This level of damage is difficult to handle with standard error correction, while here we see that these damaged packets can be used as a replacement for one missing undamaged packet. Increase of their number $d$ essentially improves the correction process. The graphs show that the Pareto assumption is not in perfect agreement with experiment, however the found coefficients describe well the general behavior.

\subsection{Summary for the general case}
Let us summarize the theoretical part: obtaining $M$ packets with correspondingly $\{\epsilon_i\}_{i=1..M}$ bit-flip probability, we should find $c$ such that $N=\sum_{i=1}^M R_c(\epsilon_i)=\sum_{i=1}^M 1-h_{1/(1+c)}(\epsilon_i)$.

As $h_u(0)=0$ and generally $h\geq 0$, if there is $N$ or more undamaged packets, no such $c$ can be found - we have the undamaged case with Gaussian distribution of weights (if receiving anything more than $N$ undamaged packets).

Otherwise, as long $c>0$, we have approximately $c$ Pareto distribution of $weight$. If $c<0$, we are below the Shannon rate: there is definitely not sufficient information for reconstruction - we have to wait for more packets. For $0\leq c \leq 1$ the (average) $weight$ for infinite packets would be infinite ($\int_0^\infty x^{-c}dx=\infty$ for $c\leq 1$). However, finite length packets can be still handled in this range. Finally, the receiver should wait until reaching some arbitrarily chosen $c_{min}$ value: \be \sum_{i=1}^M R_{c_{min}}(\epsilon_i)\geq N,\ee then perform the first trial of reconstruction. If assumed resource limit (time and memory) was not sufficient, it should wait for another packet (or more), then try reconstruction one more time and so on. The successive trials can use partial reconstruction from the previous trials.

There is a question of the minimal value ($c_{min}$) to choose before the first trial. Optimally choosing this parameter for a given scenario is a difficult question, conditioned by value of packets and available resources.
Generally $c_{min}\in[1/2,1]$ is suggested for unidirectional correction, twice smaller for bidirectional.

\section{Unknown noise level problem and concatenation with the fountain codes}     \label{ucn}
JRC allows to combine the remaining informational content of damaged packets without the need of a priory knowledge of their final damage levels. However, at least for the discussed sequential decoding, efficient joint reconstruction is rather limited to a relatively small number of packets, what can be relaxed by interleaving.

\begin{figure}[t!]
    \centering
        \includegraphics[width=8cm]{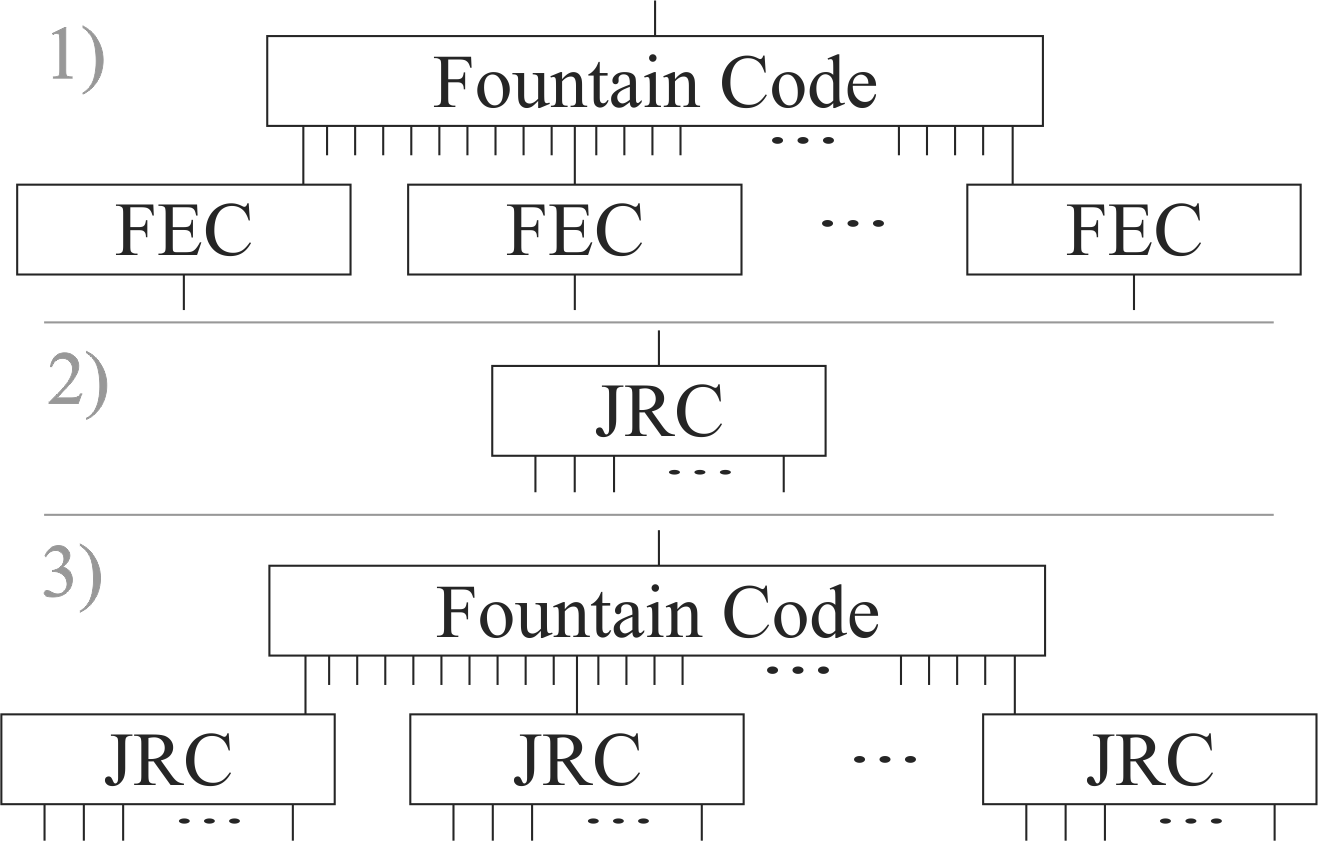}
        \caption{Three considered approaches to the unknown noise level (UNL) problem: FC+FEC, JRC, FC+JRC. While JRC can only reconstruct from a relatively small number of packets, FC can work with a very large number (we assume asymptotic setting). }
        \label{unkn}
\end{figure}

In contrast, FC, especially the Raptor codes, can easily operate on thousands of packets. They cannot reconstruct from damaged packets, what can be handled by concatenating them with FEC. However, it requires a priori knowledge of the final noise level for efficient operation. Observe that we could concatenate FC with JRC instead like in Fig. \ref{unkn}, what should allow to combine their advantages: operate on a large number of packets and with incomplete or nor a priori knowledge of the final noise level.\\

To better understand these three possibilities: FC+FEC, JRC and FC+JRC like in Fig. \ref{unkn}, let as consider an idealized problem with minimal a priori knowledge of the final noise level. For example imagine the broadcasting situation: while every receiver and packet can have some (practically random) individual noise level, the sender needs to prepare universal packets for all receivers. Let us write this problem in a general setting:

\textbf{Unknown noise level problem} (UNL): we have a large number of given type of channels, each having an independent random parameter from some probability distribution. The encoder knows the probability distribution, not the individual values. The decoder knows the parameters for all channels. How to maximize the rate?

For simplicity, let us focus on BSC channels with $\epsilon\in [0,1/2]$ bit-flip probability chosen with uniform probability distribution.\\

We would expect to approach the rate for the situation when the damage levels are known:
$$\bar{h}:=2\int_0^{1/2} \left(1-h(\epsilon)\right) d\epsilon = 1-\frac{1}{\ln(4)} \approx 0.27865$$

Let us try to handle this UNL problem using the three approaches from Fig. \ref{unkn}. For simplicity, let us assume that the used methods are perfect: FC has asymptotically $p$ rate when $p$ of packets are retrieved undamaged, FEC has rate $1-h(\epsilon)$ for BSC with $\epsilon$ bit-flip probability, JRC requires for reconstruction that sum of $1-h(\epsilon)$ over received packets is at least the number of required undamaged packets ($N$). Additionally, we are assuming here that there are no directly lost packets, however, the not fully repaired packets are treated as lost (discarded).  \\

1) \textbf{FC+FEC:} in this standard setting we need to choose redundancy level for the FECs: the largest $\epsilon$ they are able to repair. While choosing it as $\bar{\epsilon}$ for all packets, statistically $2\bar{\epsilon}$ of packets should be corrected for the assumed uniform probability $\epsilon\in [0,1/2]$. The problem of missing packets (too damaged to be repaired) is handled by the FC: the sender produces $\frac{1}{2\bar{\epsilon}}$ more packets than required for reconstruction. Finally, the best possible rate we could obtain in this approach is
$$\max_{\bar{\epsilon}\in[0,1/2]} 2\bar{\epsilon}\cdot (1-h(\bar{\epsilon})) \approx 0.11712 \quad\textrm{for}\quad \bar{\epsilon}\approx 0.15455$$
Even for the idealized conditions, this approach is still far from the optimal rate. JRC allows to improve it.\\

2) \textbf{JRC:} in pure JRC setting, we can choose two parameters: the minimal number of undamaged packets for reconstruction ($N$) and the number of actually sent packets ($M$), hence the rate is $N/M$. In the idealized situation, successful reconstruction is performed when
 $$ \sum_{i=1}^M (1-h(\epsilon_i))>N$$
We are assuming that $\epsilon_i$ are chosen randomly, hence the above condition is fulfilled only with some probability. Approximate probability of fulfilling it for $N=2$ and $M$ from 3 to 16 is succeedingly:
$$0.013, 0.063, 0.161, 0.297, 0.45, 0.593, 0.721,$$
$$0.817, 0.885, 0.931, 0.961, 0.978, 0.988, 0.994$$
for example rate $N/M=0.25$ allows to reconstruct only in $\approx 0.593$ of cases.
This inevitable problem of not correcting in all random situations was handled in the FC+FEC case by assuming asymptotic number of packet case for FC, we will use it in the next: FC+JRC setting.

Let us analyze asymptotic behavior of using pure JRC. Assuming uniform probability distribution $\epsilon\in[0,1/2]$, $X=1-h(\epsilon)$ is a random variable with $\bar{h}\approx 0.27865$ expected value and
$$\sigma=\sqrt{2\int_0^{1/2} \left(1-h(\epsilon)-\bar{h}\right)^2 d\epsilon}=\frac{\sqrt{21-2 \pi^2}}{6\ln(2)}$$
$\sigma \approx 0.26999$ standard deviation. $M$ packets are sufficient for reconstruction with probability
$$p_r(M,N)\equiv p_r=\Pr(X_1+X_2+...+X_M > N)$$
Using the central limit theorem, we get $p_r\approx 0.5$ for large $N\approx \bar{h}M$, or equivalently rate equal $N/M=\bar{h}$, what is the maximum for our problem. For a more practical $p_r$ like $0.99$, we have
$N \approx \bar{h}M - \sigma \sqrt{M}\cdot \mathrm{erf}^{-1}(p_r)$, or in other words,
$$\textrm{rate is}\quad \frac{N}{M} \approx \bar{h} - \frac{\sigma\cdot \mathrm{erf}^{-1}(p_r)}{\sqrt{M}}$$
approaching $\bar{h}$ for $M\to\infty$.\\

3) \textbf{FC+JRC} As JRC can be efficiently used for a relatively low number of packets and we would like to achieve reconstruction of all packets, we can concatenate it with FC layer to handle the problem of $p_r<1$. The FC can produce $1/p_r$ packets, leading to asymptotically complete reconstruction. Finally,
$$\textrm{rate for a given } N \textrm{ is } \max_M \frac{N}{M}\ p_r(M,N)$$
As it is shown in Table \ref{jrcfc} or can be deduced from the pure JRC case, the $\bar{h}$ rate limit is asymptotically approached. While $N\to\infty$, we rely less on FC: $p_r\to 1$.
\begin{table}[h!]
\caption{Parameters for JRC+FC and optimal choice of $M$}
\label{jrcfc}
\centering
\begin{tabular}{c|cccccccc}
  $N$ & 1 & 2 & 3 & 4  & 10 & 100  \\ \hline
  $M$ &  5 & 10 & 15 & 19  & 45 & 398 \\ 
  $p_r$ &  0.7222 & 0.8167 & 0.8715 & 0.8646  & 0.9228 & 0.9800 \\
  rate & 0.1444 & 0.1633 & 0.1743 & 0.1820 &  0.2051 & 0.2462\\
\end{tabular}
\end{table}

\section{Conclusions and further perspectives} \label{sec5}
Joint Reconstruction Codes were introduced and discussed as an enhancement of the fountain codes concept. JRC allows to combine two processes: of reconstruction from some subset of packets and of forward error correction. Standard approach in this case is using FC with packets protected by forward error correction, however, it requires a priori knowledge of damage levels, which is limited or unavailable in many scenarios. In contrast, JRC allows to operate with the same rates having only a posteriori knowledge of damage levels, at cost of being limited to a relatively small number of packets for reconstruction. It searches the space of candidates of the message to get agreement with all the received packets, tolerating disagreements accordingly to the individual damage level of each packet, which is estimated a posteriori. Thanks of it, the sender needs only to ensure a high enough statistics of packets for the receiver, do not have to care about the protection level for each individual packet - they are simultaneously payload and redundancy, decoded as having the optimal protection level. Presented theoretical analysis allows to calculate the Pareto coefficient using Renyi entropy, approximately describing the sequential decoding process.

Some examples of applications:
\begin{itemize}
\item{broadcasting: the sender cannot take individual noise levels into consideration, but needs to produce universal packets - JRC still allows to achieve rates as if every packet was optimally protected,}
\item{protecting a data storage: allowing to reconstruct the content from some a priori unknown subset of storage media, having a history dependent final damage levels,}
\item{for reducing hardware cost and energy consumption of various networks by replacing correction applied by every node, into a single reconstruction performed by the receiver,}
\item{for more robust watermarking: requiring some subset of frames or parts of a picture, with capture conditions unknown to the sender,}
\item{for communication with rapidly varying conditions, where the receiver has usually a better (a posteriori) knowledge about the actual damage level.}
\end{itemize}

The undamaged case shows that the presented approach may have also advantages as a direct replacement of FC for low number of packets. For example in the case of one excess packet ($M=N+1$), decoding needs to consider on average only twice more steps then encoding and is practically always successful. In contrast, FC often fails in this case (with $\approx 50\%$ probability).\\

This article only introduces to this new coding paradigm - there have remained plenty of research questions and possibilities for improvements and optimizations depending on specific applications.

The implemented and analyzed decoding is currently unidirectional, what could be essentially improved by using bidirectional sequential decoding, effectively doubling the Pareto coefficient~\cite{cortree}. Another line of development is including more sophisticated types of errors in the considered space of candidates/corrections - the discussed approach can even handle the synchronization errors, like deletion channel~\cite{del}. It might be also beneficial to compare the discussed sequential decoding approach with different realizations of the JRC setting, like using Low Density Parity Check~\cite{LDPC}.

Additional improvements of reconstruction efficiency can be made by optimizing the choice of transition function $f$, for example including grouping or order of the sent packets into consideration, especially when the sender is likely to receive most or all of the packets. Specifically, it should be chosen such that the set $\{extract(f[x])\}_x$ is nearly a permutation, or generally has large Hamming distances for subsets of packet which are more likely to be used for reconstruction (for example: the first sent) - increasing the probability of straightforward decoding. The choice of the bit positions for packets also influences performance (they should be far from each other). Additionally, using a Pseudorandom Number Generator initialized with a cryptographic key to choose $f$ would allow to simultaneously encrypt the message - the used key would be essential to perform reconstruction.

The main limitation of the discussed sequential decoding approach is a relatively small number of packets to reconstruct from, while FC does not have such limitation. This limitation can be relaxed by using interleaving, which efficient application requires further work. Concatenation of JRC+FC can be also beneficial, as discussed in Section \ref{ucn}.

Finally, this article assumes (a posteriori) knowledge of damage levels ($\epsilon$), what requires some estimation process. The redundancy of received damaged packets should itself allow for this purpose, suggesting that this estimation could be made as a part of decoding process. Specifically, we could start with some arbitrary $\epsilon_i$ values, then try to modify them accordingly to already decoded part. Effective exploitation and analysis of above possibilities requires further work.

\bibliographystyle{IEEEtran}
\bibliography{cites}
\end{document}